\documentclass[a4paper,11pt]{article}

\usepackage{pos}
\usepackage{lipsum} %package to generate placeholder text in the following

\usepackage{xcolor}
\usepackage{float}
\usepackage{graphicx}% Include figure files
\usepackage{dcolumn}% Align table columns on decimal point
\usepackage{bm}% bold math
\usepackage{hyperref}% add hypertext capabilities

\usepackage{lineno}
\title{Denoising radio pulses from air showers using machine-learning methods}

\ShortTitle{Denoising radio pulses from air showers using machine-learning methods}

\author*[a]{Aurélien Benoit-Lévy}
\author[b]{Zhisen Lai}
\author[b]{Oscar Macias}
\author[a,c]{Arsène Ferrière}
\onbehalf{for the GRAND Collaboration{\normalsize \\ \normalfont(a complete list of authors can be found at the end of the proceedings)}\\}

\affiliation[a]{Université Paris-Saclay, CEA, List, F-91120, Palaiseau, France}
\affiliation[b]{Department of Physics and Astronomy, San Francisco State University, 1600 Holloway
Ave, San Francisco, 94132, California, USA}
\affiliation[c]{Sorbonne Université, CNRS, Laboratoire de Physique Nucléaire et des Hautes Energies
(LPNHE), 4 Pl. Jussieu, Paris, 75005, France}

\emailAdd{aurelien.benoit-levy@cea.fr}
\emailAdd{zlai@sfsu.edu}
\abstract{

The Giant Radio Array for Neutrino Detection (GRAND) aims to detect radio signals from extensive air showers (EAS) caused by ultra-high-energy (UHE) cosmic particles. Galactic, hardware-like, and anthropogenic noise are expected to contaminate these signals. To address this problem, we propose training a supervised convolutional network known as an encoder-decoder. This network is used to learn a coded representation of the data and remove specific features from it. This denoiser is trained using high-fidelity air shower simulations specifically tailored to replicate the characteristics of signals detected by GRAND. In this contribution, we describe our machine-learning model and report initial results demonstrating the sensitivity enhancement resulting from our denoising algorithm when applied to realistically simulated GRAND signals with varying signal-to-noise ratios.
\vspace{4mm}

}

\ConferenceLogo{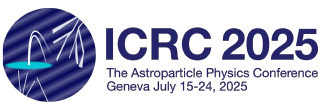}

\FullConference{39th International Cosmic Ray Conference (ICRC2025)\\  15--24 July, 2025\\ Geneva, Switzerland}

\begin{document}

\maketitle

\section{Introduction}
Radio detection experiments~\cite{Abreu_2012, ARDOUIN_2005, Tunka-Rex:2019msu, Huege_2008, LOFAR} are designed to detect the electric field from electromagnetic showers generated by cosmic particles (cosmic rays or ultra-high energy neutrinos) as they propagate through the Earth's atmosphere. While the signal from these showers can be substantial for high-energy primaries (with energies exceeding $10^7$ GeV), 
it is often obscured by a combination of noise sources. Beyond the dominant Galactic background, instrumental noise $-$ stemming from thermal fluctuations, amplifier electronics, and digitizer quantization $-$ adds a layer of uncertainty. Equally problematic is anthropogenic radio-frequency interference (RFI) (e.g., broadcast transmitters, mobile networks, and industrial equipment), which can introduce intermittent or broadband artifacts that overlap with the air-shower bandwidth.

Mitigating the combined impact of Galactic and instrumental noise requires advanced denoising strategies to lower the effective detection threshold and boost sensitivity. In recent years, machine-learning–based methods have shown great promise for disentangling these noise contributions and enhancing signal clarity. Studies have demonstrated that, when trained on realistic air shower pulses and all major noise components, deep denoisers can recover faint air-shower features with higher fidelity than traditional filtering techniques \cite{Schroder:2024Dt, Erdmann_2019} (and references therein).

Building on these advancements, this paper presents initial results demonstrating the feasibility of using machine learning techniques to efficiently denoise the traces obtained through the data acquisition system of the Giant Radio Array for Neutrino Detection (GRAND) \cite{Alvarez_Muniz_2019}  prototype detectors, GRANDProto300. Our approach uses state-of-the-art algorithms to distinguish between the desired signals and the pervasive background noise, thereby enhancing the overall effectiveness of radio detection experiments. This contribution not only showcases the potential of machine learning in this domain but also paves the way for future innovations in the detection and analysis of high-energy cosmic rays and neutrinos.

\section{Methodology}

In this work, we investigate a purely data-driven approach in which we use only data, whether simulated or real, and make no other assumptions about physical models or theories. The only principle we follow is that, since this denoising algorithm ultimately aims to be used on real data, we must ensure that the training data is as close as possible to the real data. In this section, we present the architecture of the model we considered and describe the construction of the training dataset.

\subsection{Denoiser Model}

Extensively studied in computer vision, the task of denoising is the simplest of all signal reconstruction tasks. Much, if not all, of the machinery developed for image reconstruction can be used for the processing of radio traces. Radio traces with 2 or 3 polarizations can be seen as 1D images that have just 1 pixel height and 2 or 3 channels. We thus consider an encoder-decoder architecture. 
Specifically, the encoder compresses the input time series into a concise latent representation, while the decoder reconstructs the original signal from this compressed form. While the overall architecture of our model is generic, we introduce two novelties that aid the training procedure and/or provide better results. First, for both the encoder and the decoder, we use residual blocks \cite{7780459}, which are a key ingredient in all recent neural network architectures. The residual blocks enhance feature extraction by allowing gradients to flow through shortcut connections, which helps to train deeper networks without the risk of vanishing gradients and improves the model's ability to learn complex patterns. Second, we divide the encoder part into two branches: one in the time domain and the other in the frequency domain. Each serves a unique purpose in feature extraction.
The time domain branch is designed to capture local patterns and temporal dependencies in the time-series data using one convolutional layer followed by two residual blocks.
The frequency domain branch is designed to obtain the frequency representation and periodic patterns of the ADC signal traces by applying a Fast Fourier Transform (FFT) to the input signals. As a result, the FFT transforms 3 channels of input signals into 3 imaginary parts and 3 real parts of frequency components, which are then stacked and fed to one convolutional layer followed by two residual blocks, as in the time domain branch.
It should be noted that the model is fully convolutional \cite{FCN}, allowing it to be used on traces of arbitrary length.

\begin{figure*}[t!]
    \centering
    \includegraphics[width=1\linewidth]{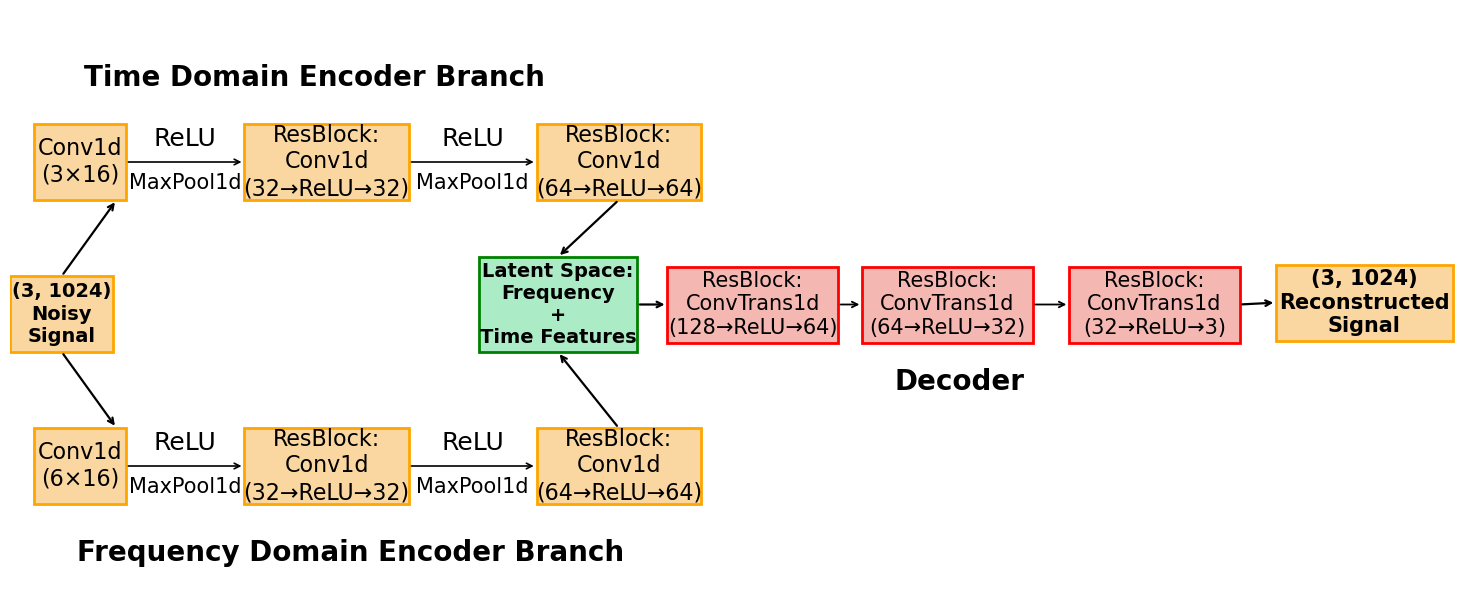}
    \caption{Architecture of our encoder-decoder architecture. The encoder part (yellow blocks) is composed of a time domain and a frequency domain branches, while the decoder part (red blocks) reconstructs the input signal to produce a denoised trace with the same dimension as the noisy input. 
    \label{fig:enter-label}}
\end{figure*}

\section{Simulations and data}
\subsubsection*{Cosmic ray radio signals}
For the radio signal generated by cosmic ray (CR) showers, we used 8000 events simulated with the ZHAireS package \cite{Alvarez_Mu_iz_2012}. ZHAireS is a well-established tool used to simulate the electric fields from radio emissions produced by extensive air showers. These electric fields are then processed using the GRAND internal software, GRANDLib \cite{2025109461}, to produce voltage traces similar to those produced by the GRANDProto300 detectors \cite{grand, gp300}, taking into account electronics and the full RF-chain.
Out of the many events that are simulated, many antennas produce very weak signals. In order to have a balanced dataset, we only retain traces for which the maximum amplitude in one of the two polarizations is above 15 ADC, ensuring that we focus on significant signals for our analysis.
Our training set contains 138,769 pairs of traces, while the validation and test sets each have 21,145 pairs of traces.
\subsubsection*{Noise}
In our work on noise removal techniques, the construction of training and testing noise sets is of paramount importance. The primary sources of homogeneous noise include the Galactic radio background and thermal noise. We focus exclusively on homogeneous noise, as opposed to transient noise, which can originate from anthropogenic sources that are often challenging to identify and simulate. It should also be noted that the separation between astrophysical and anthropogenic pulses is better done with a classifying approach \cite{lecoz} rather than with denoising techniques. 

\begin{figure}[ht]
    \centering
    \includegraphics[width=1.0\linewidth]{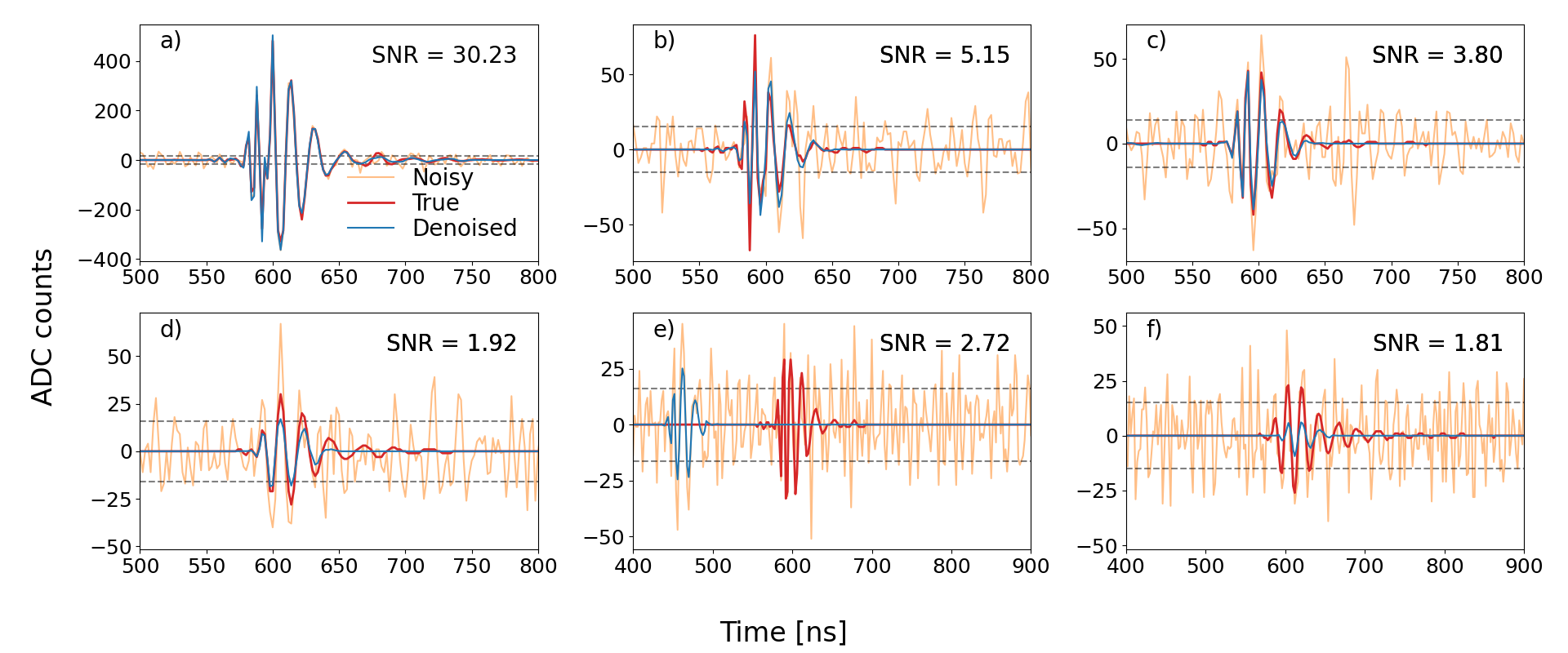}
    \caption{Examples of denoised traces for varying signal-to-noise ratios (SNRs). In all panels, the raw or noisy traces are shown as orange lines, the true signal that we aim to recover is depicted in red, and the denoised traces are illustrated in blue. The dashed lines represent the $1\sigma$-level of the noise in the corresponding trace. Panels a) to d) present examples where the denoising procedure was successful in terms of both denoised amplitude and peak position. Panel e) shows an example where the denoising is considered successful with respect to the two conditions described in Section~\ref{sec_results}, but the recovered peak time is not at the correct position. Panel f) shows an example where the denoised amplitude is too low.}
    \label{fig_examples}
\end{figure}
To address these challenges, a common approach involves using real measured noise traces to train machine learning models on mixed datasets comprising simulated signals and real noise. However, our method diverges slightly: we utilize real noise traces solely to extract their mean power spectra. This allows us to generate Gaussian noise that mirrors the power spectra of the actual data. Specifically, we employ the ADC noise (AN) traces detailed in the NUTRIG proceedings~\cite{nutrig}. These traces are grouped by Data Unit (DU), and their power spectra are averaged over brief intervals, typically spanning a few tens of minutes. Consequently, we obtain a collection of noise power spectra that accurately represent the noise characteristics of real data. We then randomly select these power spectra to produce random noise traces, which, while Gaussian, possess realistic frequency content.

\begin{figure}[H]
    \centering
    \includegraphics[width=0.95\linewidth]{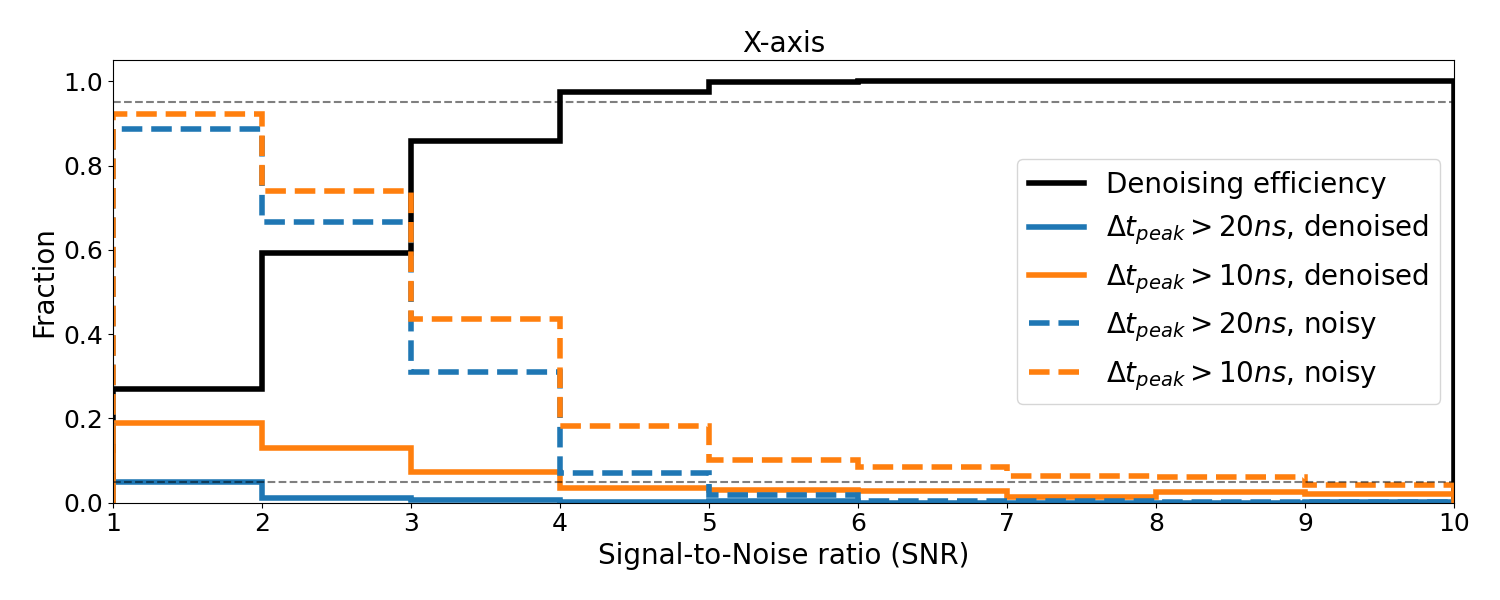}
    \caption{Denoising efficiency (solid black line) for an ADC threshold of 15 ADC. The gray dashed lines represent the 5\% and 95\% levels. The colored lines represent the fraction of denoised (solid lines) or noisy (dashed lines) traces that have their peak time off by more than 10 ns (orange lines) or 20 ns (blue lines) compared to the peak time of the clean signal. South-North axis.}
    \label{fig_correct_fractionX}
\end{figure}
\begin{figure}[ht]
    \centering
    \includegraphics[width=0.95\linewidth]{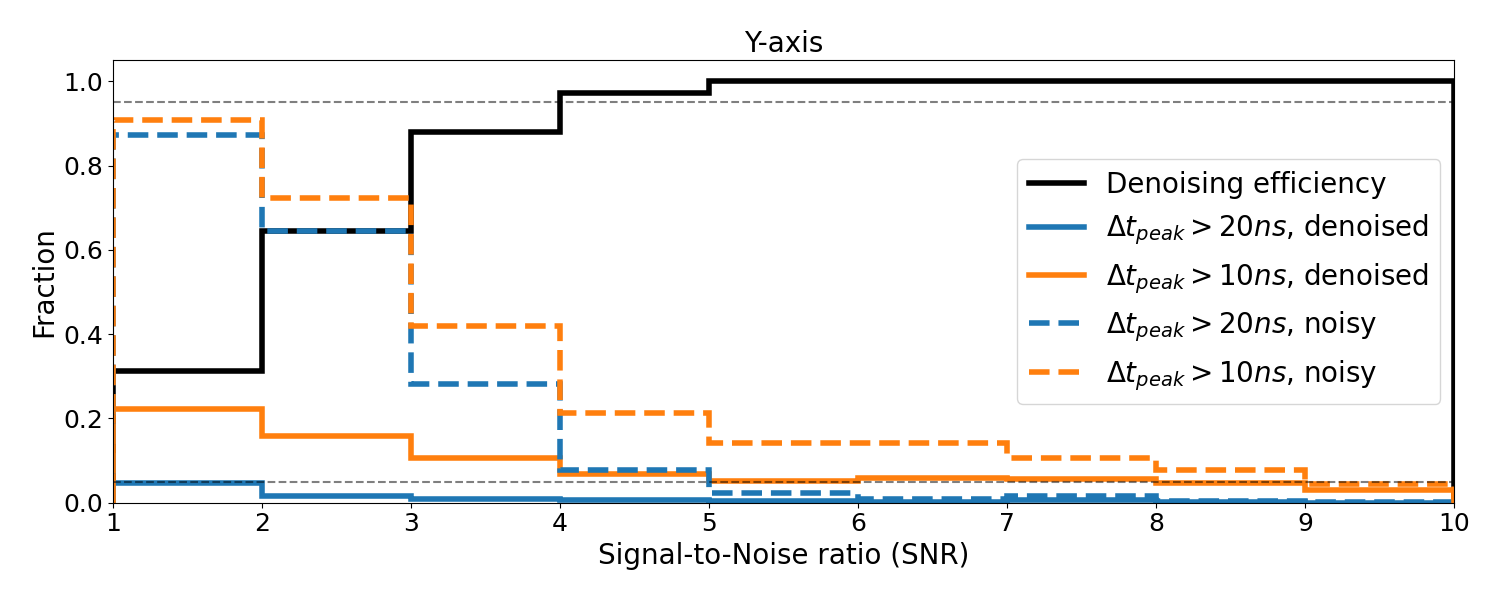}
    \caption{Same as Figure \ref{fig_correct_fractionX} for the East-West axis.}
    \label{fig_correct_fractionY}
\end{figure}

Another rationale for excluding real noise traces from the training set is the potential presence of low signal-to-noise ratio (SNR) signals within those traces. Training a model with such hidden signals could prove counterproductive. Additionally, one of the inherent challenges in applying machine learning to physical sciences is bridging the reality gap—ensuring that models trained on simulations perform effectively with real-world data. As we will illustrate in Section~\ref{sec_results}, our approach successfully meets this challenge, thereby significantly bolstering its validity.

Finally, the noisy traces are constructed by combining simulated signals with simulated noise traces. Due to the construction of the ZHAireS traces, the pulse consistently appears at the same position within the traces (around time bin 300). To prevent the network from learning this specific position, we avoid training on the full 1024 time bin traces. Instead, we use a randomly selected 768 time bin window. Since we train and test the machine learning models with traces constructed from separate signals and noise, we have access to the noiseless signal. We therefore define the Signal-to-Noise Ratio (SNR) of a noisy trace $X_i$ that contains a signal as the maximum of the Hilbert envelope of the noiseless trace $\hat{X_i}$ divided by the standard deviation of the region of the trace that does not contain the signal:
\begin{equation}
\text{SNR} = \frac{\max_i |H(\hat{X_i})|}{\text{std}(X_i)}
\end{equation}

\section{Training procedure}
The training procedure is fairly standard. We use the Adam \cite{adam} optimizer with a weight decay of \(5 \times 10^{-4}\). We use a cyclic learning rate \cite{cyclicLR}, with mode \texttt{triangular2}, with a base learning rate of \(1 \times 10^{-6}\), a maximum learning rate of \(3 \times 10^{-5}\), and a step size chosen so that the learning rate undergoes 10 full cycles across 600 epochs and a batch size of 64. In practice, all these hyperparameters matter very little in the final results. The model is trained by minimizing the \(L_1\) loss between the noiseless trace $\hat{X}$ and its reconstruction $\bar{X}$:
\begin{equation}
L = \frac{1}{N}\sum_{t=1}^{N}|\hat{X_t} - \bar{X}_t|.
\end{equation}
The \(L_1\) loss was chosen as it significantly decreases the rate of false reconstruction, compared to more classical losses like the Mean-Squared Error (MSE, or \(L_2\) loss).

\section{Results\label{sec_results}}
While the noise part of the traces we used for training is simulated Gaussian noise, we used the real noise traces from the AN noise dataset~\cite{nutrig} for the testing set and the determination of the performances of the model. A visual inspection of the denoised traces of the test set indicate good general performances, and we present a few examples of successfully denoised traces in Figure~\ref{fig_examples}.
In order to estimate the performance of our denoising model, and as is standard practice in machine learning studies, we first estimate the number of false positives and false negatives. Indeed, we are interested in knowing how well the denoiser actually finds the signals that are hidden in the noisy traces. So, we want to minimize the quantity of false negatives. Similarly, we are interested in quantifying the proportion of signals found by the denoiser that correspond to real signals. Hence, we want to minimize the number of false positives. Given a noisy trace, we consider it successfully denoised if the following conditions are met:
\begin{description}
    \item[Condition 1:] The peak amplitude of the clean trace is above the chosen ADC threshold.
    \item[Condition 2:] The peak amplitude of the denoised trace is above the chosen ADC threshold.
\end{description}

Given these conditions, we define the denoising efficiency as the ratio of the number of traces satisfying conditions 1 and 2 to the number of traces satisfying condition 1. This denoising efficiency is presented (black solid line) in Figures \ref{fig_correct_fractionX} and \ref{fig_correct_fractionY} for the South-North and East-West axes, respectively, for an ADC threshold of 15 ADC. As can be seen, the denoising efficiency is above 95\% for signals with SNR $\approx$ 4.
Another important quantitative test is the position of the peak of maximum amplitude. For the traces that are successfully denoised (meeting conditions 1 and 2 above), we compute the time difference $\Delta t_{\text{peak}}$ between the time of the maximum peak amplitude for the denoised trace and the time of the maximum peak amplitude for the noiseless trace. We also compute this quantity for the noisy traces. In that case, and for low SNR traces, the peak maximum is likely to correspond to a high noise value which would be evenly distributed with the trace. In Figures \ref{fig_correct_fractionX} and \ref{fig_correct_fractionY}, we show the fraction of traces for which $\Delta t_{\text{peak}} > 10$ ns (orange lines) and $\Delta t_{\text{peak}} > 20$ ns (blue lines), for the denoised traces (solid lines) and the noisy traces (dashed lines). These thresholds were chosen as they are of the same magnitude as the precision of the GPS timings of the GRANDProto300 detectors~\cite{gp300}. An example of a denoised trace which has a time difference larger than these thresholds is shown in the e) panel of Figure~\ref{fig_examples}.

As can be seen, denoising is efficient for intermediate SNRs (SNR $\approx$ 4-5), as the fraction of incorrect peak positions is drastically reduced. It is also effective for high SNR traces (SNR $> 5$), for which a small but non-negligible fraction (around 5\%) have their peak position wrong by more than 10 ns.

\begin{figure}[ht]
\begin{minipage}[t]{0.48\linewidth}
\begin{center}
\includegraphics[width=0.99\linewidth]{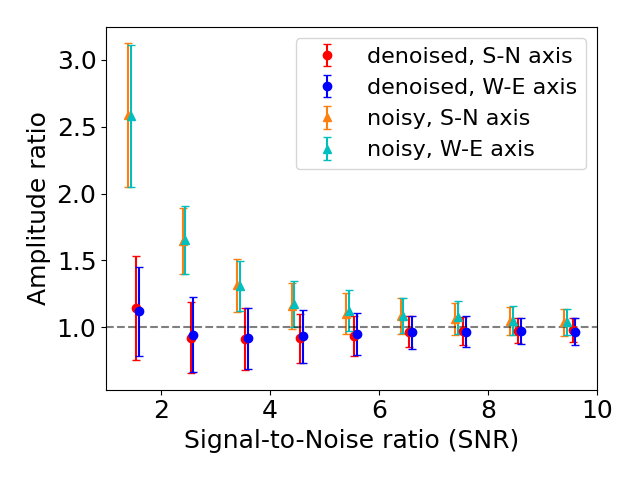}
\end{center}
\end{minipage}
\begin{minipage}[t]{0.48\linewidth}
\begin{center}
\includegraphics[width=0.99\linewidth]{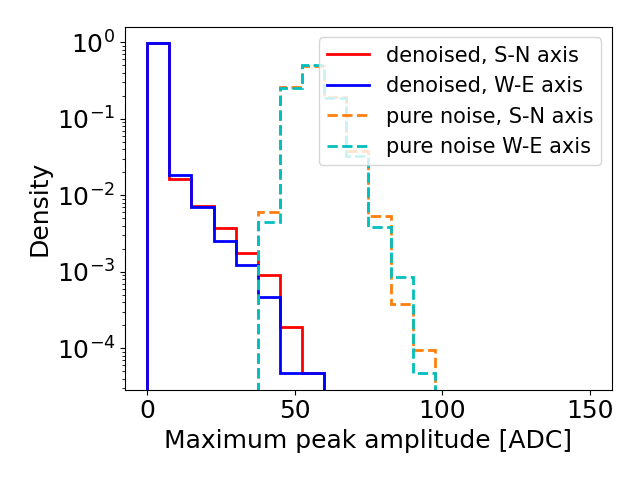}
\end{center}
\end{minipage}
\caption{\label{amp_ratio}{\bf Left:} Ratio of the peak amplitude of successfully denoised (red and blue dots) and noisy (orange and cyan triangles) traces to the peak amplitude of the noise-free signals. Denoising efficiently enables the recovery of the real signal for low SNR traces. {\bf Right:} Distribution of the maximum peak amplitude for traces containing only noise (orange and cyan dashed-lines), and the resulting distribution after denoising (red and blue solid lines).}
\end{figure}

We show in Figure \ref{amp_ratio} (left panel) the ratio between the maximum peak amplitude of the successfully denoised (red and blue dots) or noisy (orange and cyan triangles) traces and the maximum peak amplitude of the clean traces. As expected, as the SNR decreases, this ratio diverges for the noisy traces, while it remains compatible with one in the denoised case.

Finally, we measure the false positive fraction. For this test, we ran the denoiser on a full test set composed of pure Gaussian AN noise. In Figure \ref{amp_ratio} (right panel), we present the distribution of the maximum peak amplitude for the denoised traces (red and blue solid lines) and for the noisy traces (orange and cyan dashed lines). The vast majority of the noise traces have their maximum peak amplitude below 10 ADC counts, and the fraction of false positives above an ADC threshold of 15 (which roughly corresponds to SNR = 1) is 1.4\% for the S-N channel and 1.1\% for the E-W channel. If we increase this threshold to 40 (SNR$\sim 3$), this fraction vanishes. Moreover, the few noise traces that exhibit very high amplitude are successfully denoised.

\section{Conclusion}
In this study, we presented a robust approach to denoising radio traces from the GRANDProto300 experiment using a machine learning model, specifically an encoder-decoder architecture. By using simulated data and realistic noise conditions, we demonstrated the effectiveness of our method in accurately recovering signals even at low signal-to-noise ratios. Our approach enables successful denoising of traces at the 95\% level for SNR > 4, while maintaining the false positive rate under control and preserving the integrity of the signal's peak position and amplitude. Furthermore, the fact that the model has been trained with simulated Gaussian noise and successfully applied to real noise traces from the experiment underscores its potential for real-world deployment. These results not only demonstrate the effectiveness of our method but also pave the way for its integration into ongoing and future experiments in radio detection of cosmic rays and ultra-high energy neutrinos.

% Bibtex references:
\bibliographystyle{ICRC}
\bibliography{references}
% Alternatively, you can include references by hand:
%\begin{thebibliography}{99}
%\bibitem{...}
%
%\end{thebibliography}

\section*{Full Author List: GRAND Collaboration}

\scriptsize
\noindent
J.~Álvarez-Muñiz$^{1}$, R.~Alves Batista$^{2, 3}$, A.~Benoit-Lévy$^{4}$, T.~Bister$^{5, 6}$, M.~Bohacova$^{7}$, M.~Bustamante$^{8}$, W.~Carvalho$^{9}$, Y.~Chen$^{10, 11}$, L.~Cheng$^{12}$, S.~Chiche$^{13}$
% ...
\\
\\
$^{1}$Departamento de Física de Particulas \& Instituto Galego de Física de Altas Enerxías, Universidad de Santiago de Compostela, 15782 Santiago de Compostela, Spain \\
$^{2}$Institut d'Astrophysique de Paris, CNRS  UMR 7095, Sorbonne Université, 98 bis bd Arago 75014, Paris, France \\
$^{3}$Sorbonne Université, Université Paris Diderot, Sorbonne Paris Cité, CNRS, Laboratoire de Physique  Nucléaire et de Hautes Energies (LPNHE), 4 Place Jussieu, F-75252, Paris Cedex 5, France \\
$^{4}$Université Paris-Saclay, CEA, List,  F-91120 Palaiseau, France \\
$^{5}$Institute for Mathematics, Astrophysics and Particle Physics, Radboud Universiteit, Nijmegen, the Netherlands \\
$^{6}$Nikhef, National Institute for Subatomic Physics, Amsterdam, the Netherlands \\
$^{7}$Institute of Physics of the Czech Academy of Sciences, Na Slovance 1999/2, 182 00 Prague 8, Czechia \\
$^{8}$Niels Bohr International Academy, Niels Bohr Institute, University of Copenhagen, 2100 Copenhagen, Denmark \\
$^{9}$Faculty of Physics, University of Warsaw, Pasteura 5, 02-093 Warsaw, Poland \\
$^{10}$Key Laboratory of Dark Matter and Space Astronomy, Purple Mountain Observatory, Chinese Academy of Sciences, 210023 Nanjing, Jiangsu, China \\
$^{11}$School of Astronomy and Space Science, University of Science and Technology of China, 230026 Hefei Anhui, China \\
$^{12}$National Astronomical Observatories, Chinese Academy of Sciences, Beijing 100101, China \\
$^{13}$Inter-University Institute For High Energies (IIHE), Université libre de Bruxelles (ULB), Boulevard du Triomphe 2, 1050 Brussels, Belgium \\
% ...

%%%%%%%%%%%%%%%%%%%%%%%%%%%%%%%%%%%%%%%%%%%%%%%%%%%%%%%%%%%%%%
%%%%%%%%%%%%%%%%%%%%%%%%%%%%%%%%%%%%%%%%%%%%%%%%%%%%%%%%%%%%%%

\subsection*{Acknowledgments}

\noindent
The GRAND Collaboration is grateful to the local government of Dunhuag during site survey and deployment approval, to Tang Yu for his help on-site at the GRANDProto300 site, and to the Pierre Auger Collaboration, in particular, to the staff in Malarg\"ue, for the warm welcome and continuing support.
The GRAND Collaboration acknowledges the support from the following funding agencies and grants.
%%%%
\textbf{Brazil}: Conselho Nacional de Desenvolvimento Cienti\'ifico e Tecnol\'ogico (CNPq); Funda\c{c}ão de Amparo \`a Pesquisa do Estado de Rio de Janeiro (FAPERJ); Coordena\c{c}ão Aperfei\c{c}oamento de Pessoal de N\'ivel Superior (CAPES).
%%%%
\textbf{China}: National Natural Science Foundation (grant no.~12273114); NAOC, National SKA Program of China (grant no.~2020SKA0110200); Project for Young Scientists in Basic Research of Chinese Academy of Sciences (no.~YSBR-061); Program for Innovative Talents and Entrepreneurs in Jiangsu, and High-end Foreign Expert Introduction Program in China (no.~G2023061006L); China Scholarship Council (no.~202306010363); and special funding from Purple Mountain Observatory.
%%%%
\textbf{Denmark}: Villum Fonden (project no.~29388).
%%%%
\textbf{France}: ``Emergences'' Programme of Sorbonne Universit\'e; France-China Particle Physics Laboratory; Programme National des Hautes Energies of INSU; for IAP---Agence Nationale de la Recherche (``APACHE'' ANR-16-CE31-0001, ``NUTRIG'' ANR-21-CE31-0025, ANR-23-CPJ1-0103-01), CNRS Programme IEA Argentine (``ASTRONU'', 303475), CNRS Programme Blanc MITI (``GRAND'' 2023.1 268448), CNRS Programme AMORCE (``GRAND'' 258540); Fulbright-France Programme; IAP+LPNHE---Programme National des Hautes Energies of CNRS/INSU with INP and IN2P3, co-funded by CEA and CNES; IAP+LPNHE+KIT---NuTRIG project, Agence Nationale de la Recherche (ANR-21-CE31-0025); IAP+VUB: PHC TOURNESOL programme 48705Z. 
%%%%
\textbf{Germany}: NuTRIG project, Deutsche Forschungsgemeinschaft (DFG, Projektnummer 490843803); Helmholtz—OCPC Postdoc-Program.
%%%%
\textbf{Poland}: Polish National Agency for Academic Exchange within Polish Returns Program no.~PPN/PPO/2020/1/00024/U/00001,174; National Science Centre Poland for NCN OPUS grant no.~2022/45/B/ST2/0288.
%%%%
\textbf{USA}: U.S. National Science Foundation under Grant No.~2418730.
%%%
Computer simulations were performed using computing resources at the CCIN2P3 Computing Centre (Lyon/Villeurbanne, France), partnership between CNRS/IN2P3 and CEA/DSM/Irfu, and computing resources supported by the Chinese Academy of Sciences.

\end{document}